\documentstyle[prl,twocolumn,aps,epsfig]{revtex}
\begin{document}
\draft
\title{Superfluid and Dissipative Dynamics of a Bose-Einstein
  Condensate in a Periodic Optical Potential}
\author{S. Burger, F. S. Cataliotti, C. Fort, F. Minardi, and M. Inguscio}
\address{INFM - LENS, Dipartimento di Fisica,
 Universit\`a di Firenze, Largo Enrico Fermi 2, I-50125 Firenze, Italia}
\author{M. L. Chiofalo and M. P. Tosi}
\address{INFM - Scuola Normale Superiore, 
Piazza dei Cavalieri 7, I-56126 Pisa, Italia}
\date{\today}
\maketitle
\begin{abstract}
We create Bose-Einstein condensates of $^{87}$Rb in a static
magnetic trap with a superimposed blue-detuned 1D optical lattice.
By displacing the magnetic trap center we are able to control the
condensate evolution. We observe a change in the frequency of the
center-of-mass oscillation in the harmonic trapping potential, in
analogy with an increase in effective mass. For fluid
velocities greater than a local speed of sound, we observe
the onset of dissipative processes up to full removal of the
superfluid component. A parallel simulation study visualizes the
dynamics of the BEC and accounts 
for the main features of the observed behavior.
\end{abstract}
\medskip

%03.75.Fi Phase coherent atomic ensembles; quantum condensation phenomena
%32.80.Pj Optical cooling of atoms; trapping
%32.80.Qk Coherent control of atomic interactions with photons
%67.57.De Superflow and hydrodynamics

\narrowtext

Bose-Einstein condensates (BEC) in dilute atomic gases are
macroscopic quantum systems which can be  manipulated  by a
variety of experimental techniques~\cite{Varenna}. The current
development of such techniques is opening up a wealth of
possibilities to explore new physics, e.g., in non-linear atom
optics~\cite{Deng1999a}, and to study various aspects of superfluid 
behavior in the precisely controllable  context of atomic physics
~\cite{Raman1999a}.

Atoms confined in a periodic potential share some properties
with systems of electrons in crystals. Effects known from
solid state physics, like Bloch oscillations and
Wannier-Stark ladders, have been observed by exposing cold atoms to
the dipole potential of far detuned optical
lattices~\cite{Dahan1996a}. Macroscopic quantum interference has been
observed in an experiment on 
a BEC confined to the antinodes of a far detuned optical
lattice~\cite{Anderson1998a}. Bragg diffraction from a
condensate has been induced in moving optical
lattices~\cite{Kozuma1999a}. This has been used, e.g., as an
atom-laser outcoupler~\cite{Hagley1999a} and as a tool for spectroscopy
of the momentum in BEC's~\cite{Stenger1999b}. 
Applications of BEC's
in periodic potentials range from matter-wave
transport~\cite{Zobay1999a} to
interferometry~\cite{Anderson1998a} and  quantum
computing~\cite{Jaksch1998b}. 
The question of the stability of the BEC during the
evolution in optical potentials is crucial for these applications
and has been addressed in theoretical
works~\cite{Bronski2000x}.

In this Letter we report on some novel aspects of superfluidity in
BEC's by studying their center-of-mass oscillations inside
the harmonic potential of a magnetic trap in presence of a
one-dimensional (1D) optical lattice. We identify different
dynamical regimes by varying the initial displacement of the BEC from
the bottom of the trap.
For small displacements the BEC performs undamped oscillations in the
harmonic potential and feels the periodic potential only
through a shift in the oscillation frequency.
At larger displacements we observe the onset of dissipative processes
appearing through a damping in the oscillations. We can describe the
experimental results in terms of an inhomogeneous superfluid having a
density-dependent critical velocity.
In parallel we report numerical studies 
of the Gross-Pitaevskii equation (GPE), which
capture the main features of the observed dynamics.

In our experimental setup~\cite{Fort2000a} we now produce BEC's 
of $^{87}$Rb atoms in the (F=1,m$_F$=\,$-1$) state. 
The fundamental frequencies of our Ioffe-type magnetic trap are
$\omega_x=2\pi\times 8.7$\,Hz and $\omega_{\perp}=2\pi\times90\,$Hz 
along the axial and radial directions, respectively. 
The condensates are cigar-shaped with the long axis (the $x$-axis)
oriented horizontally. With a number of atoms $N=4\times 10^5$, the
typical dimensions (Thomas-Fermi radii)
are $R_x=55\,\mu$m and $R_\perp=5.5\,\mu$m.

We create a 1D optical lattice by superimposing to the long axis
of the magnetic trap a far detuned,
retroreflected laser beam with wavelength $\lambda$. 
The waist of the beam is two orders of
magnitude larger than the short condensate axis, and therefore the
resulting dipole potential in the condensate region has the form
$V(\vec{r}\,) = V_0\, \cos^2(2\pi x/\lambda)$. 
With a blue
detuning $\delta=2\pi\times 50\,$GHz from the D1-line at
$\lambda_0\simeq 795\,$nm and an intensity $I=1\,$mW/mm$^2$ in
the antinodes of the standing wave, the dipole potential height
of an optical lattice well is $V_0/k_B \simeq  270\,$nK~\cite{Note1}. 
The
spontaneous scattering rate in the 
antinodes is $\Gamma_{sp}\simeq 0.7\,$Hz
at this detuning and intensity. 

To prepare the atomic cloud in the ground state of
the combined magnetic trap and optical lattice we first perform
evaporative cooling in the magnetic trap until we reach a
temperature slightly above the critical temperature, $T\simeq
1.5\,T_c$. Then we superimpose the optical lattice and continue
with the evaporation process down to a temperature $T\simeq T_c/2$,
where the thermal cloud is no longer observable. We have checked
that the time at which we switch on the optical lattice does only
affect the atom number according to the spontaneous scattering
rate, but does not influence the BEC dynamics as long as the
potential is switched on at temperature above $T_c$. It is
important to note that, for the dipole potential strengths and
atom numbers which are used in the experiments, the modulated atomic
density does not vanish in the antinodes of the lattice but reaches
a minimum value which is significantly
different from zero.

\begin{figure}
   \begin{center}
   \parbox{6cm}{
   \epsfxsize 6cm
   \epsfbox{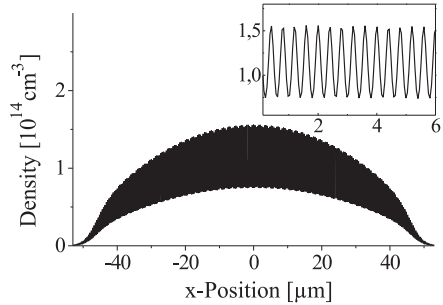}
}
   \end{center}
\caption {\small
Density distribution of a BEC in a harmonic trap with a superimposed optical
lattice, from a numerical simulation of the 3D GPE for
$N=3\times10^5$ and $V_0/k_B=270\,$nK.
The inset shows an enlargement of the central region of the BEC.
The envelope of the modulated density distribution follows
the parabolic distribution in the harmonic trap.
}
\label{simulation_groundstate}
\end{figure}

Figure~\ref{simulation_groundstate} shows the density
distribution along the $x$-axis, as obtained by numerical
propagation of the 3D GPE in 
imaginary time \cite{Chiofalo2000a}. 
Here, the
condensate spans about 250 lattice sites. In the experiment, the
density  modulation on the length scale of $\lambda/2$ cannot be
resolved, due to the limited resolution ($\simeq 7\,\mu$m)
of the absorption-imaging system. The
modulation on a short length scale raises the chemical
potential to the value $\mu/k_B\simeq 170\,nK$  for 
$N=3\times 10^5$ in the
combined trap. Instead, in the purely
magnetic trap we have $\mu/k_B\simeq 47\,nK$. 

In order to investigate the dynamics of the system we translate
the magnetic trapping potential in the $x$-direction by a
variable distance $\Delta x$ ranging up to $300\,\mu$m 
by changing the currents
through the coils of the magnetic trap.
The translation takes a 
few milliseconds, which is short compared to the
longitudinal oscillation period $2\pi/\omega_x$. Therefore,
the BEC finds itself out of equilibrium and is subject to a
potential gradient which forces it into motion. 
The presence of the
magnetic trapping potential  ensures that the atomic cloud maintains 
its high density (maximum density $n_{max}\approx
1.5\times 10^{14}\mbox{cm}^{-3}$). After an evolution time $t_{ev}$
in the displaced trap, both the magnetic trapping and
the optical lattice are switched off simultaneously and the
cloud is imaged after an additional free expansion of 26.5\,ms.
The imaging beam is horizontal and directed perpendicularly to
the long condensate axis. From the absorption image we
deduce the center-of-mass motion and gain information on the
distortion of the BEC.

In the absence of the optical lattice,
the center-of-mass motion of the BEC
in the displaced trap is
an undamped oscillation with frequency $\omega_x=2\pi\times
8.7\,$Hz and amplitude $\Delta x$, to which in the following we
refer to as the ``free oscillation''. After switching on the optical 
lattice we observe dynamics in different regimes.

\begin{figure}
   \begin{center}
   \parbox{5.2cm}{
   \epsfxsize 5.2cm
   \epsfbox{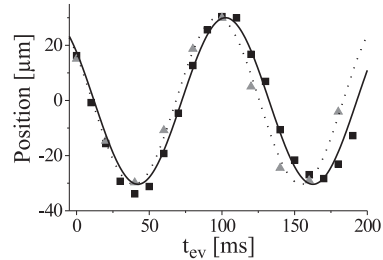}
}
   \end{center}
\caption {\small 
Superfluid oscillations of a BEC in the presence of an optical lattice
potential of height $V_0/k_B\simeq 270\,$nK (squares)
and in a purely magnetic trap (triangles), for  
initial displacement $\Delta x=(31\pm3)\,\mu$m.
The lines give results from a numerical simulation of the 
1D GPE at the experimental parameters.
}        
\label{super_oszi}
%Data from 20.12.2000 quad ap 93 and 650microW lattice power
%Data from 20.12.2000 quad ap 94 and 650microW lattice power
\end{figure}

For small displacements, $\Delta x<\approx 50\,\mu$m, the
dynamics of the BEC resembles the free oscillation at the same
amplitude but with a significant shift in frequency.
Figure~\ref{super_oszi} shows a comparison of free oscillations
and oscillations with superimposed lattice for $\Delta
x=(31 \pm 3)\,\mu$m. For the lattice potential $V_0/k_B\simeq 270\,$nK
we find a shifted frequency $\omega^\ast =2\pi\times (8.0\pm
0.1)\,$Hz. As can be seen from Fig.~\ref{super_oszi},
this frequency shift is also reproduced in numerical
simulations of the 1D GPE using an explicit
time-marching method~\cite{Chiofalo2000a,Note2}. 

The frequency shift can be explained in 
terms of a renormalization of the atomic mass
in the band states originating from the periodic potential. From the data in 
Fig.~\ref{super_oszi} we obtain an effective mass 
$m^\ast/m=(\omega_x/\omega^\ast)^2=1.18\pm 0.02$. Different from 
earlier experiments on cold atoms in an optical lattice under constant
acceleration~\cite{Dahan1996a}, in the present small-amplitude regime 
under harmonic forces we are only 
exploring the states near the Brillouin zone  
center. The above value of $m^\ast$ refers, 
therefore, to states near the bottom
of the energy band.

The essentially undamped oscillations of the BEC on the time scale of the
experiments in the present small-amplitude regime is a manifestation of
superfluid behavior. The coherent condensate is being accelerated through band
states as if it were a quasi-particle~\cite{Chiofalo2000x}. 
Also, for small displacements 
we observe only marginal heating effects, i.e., the small thermal
cloud of atoms accompanying the BEC can be fully accounted for by
spontaneous scattering. %perhaps shorten

\begin{figure}
   \begin{center}
   \parbox{8cm}{
   \epsfxsize 8cm
   \epsfbox{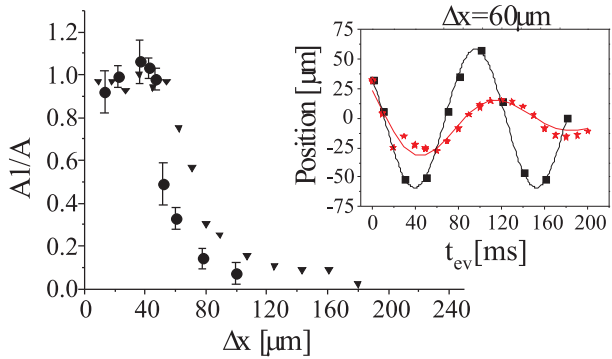}
}
   \end{center}
\caption {\small 
Ratio of the first-peak amplitude of the oscillation of the ensemble to 
the free-oscillation amplitude, $A_1/A$, as a function of initial displacement 
$\Delta x$,
for the potential $V_0/k_B\simeq 270\,$nK and atom number 
$N\simeq 3\times 10^5$.
Circles: experimental data, triangles: results from 1D numerical simulation.
The inset shows a full oscillation for a displacement $\Delta x=60\,\mu$m
with (stars) and without (squares) optical lattice.
Here, lines are fits to the data.
}        
\label{smorz}
\end{figure}

However, the BEC enters a regime of dissipative dynamics when
we further increase the initial 
displacement $\Delta x$ and hence the velocity of the
condensate. 
In Fig.~\ref{smorz} we report the measured ratio between
the first oscillation peak amplitude and the free-oscillation 
amplitude as a function of the 
trap displacement, together with the values from the
numerical simulation.
At a displacement $\Delta x\simeq 50\,\mu$m when the maximum
velocity attained by the condensate is $v\simeq 3\,$mm/s, 
this ratio suddenly deviates from unity, indicating the
insurgence of dissipation in the condensate motion. As shown in
the inset of Fig. \ref{smorz}, the subsequent dynamics is a damped
oscillation of the center-of-mass at a greatly reduced frequency.
The damping increases by further increasing 
the initial displacement, as is
seen in the main body of Fig. \ref{smorz}, 
this behavior being also displayed
by the simulation data in the same Figure.

With the onset of dissipative processes the condensate
shape in the experiment becomes distorted and a much broader distribution,
compatible with
a thermal component appears.
A thermal component is not allowed to arise in the numerical simulation, 
which is still based on the GPE. However, the 
density distribution of the condensate in the simulation becomes fragmented and 
its phase is completely randomized. That is, the condensate in this regime 
breaks up into subsystems residing in an essentially independent manner inside 
the various wells of the periodic potential.

Superfluidity can be expected to disappear when
the velocity of flow is sufficient for the spontaneous emission of elementary 
excitations, as is the case for a homogeneous Bose gas~\cite{Nozieres}. 
In a trapped condensate moving at sufficiently high velocities, 
emission of phonons and other excitations is favoured and the gas
becomes heated~\cite{Raman1999a,Kagan2000b}. 
The essentially 1D dynamics of the present
sample implies an important role for longitudinal phonon excitations 
in these processes, with a spectrum of critical 
velocities because of the inhomogeneity.
In a simplified picture,
the optical lattice can be viewed as a medium with a
microscopic roughness, which leads to a velocity-dependent local
compression of the gas moving through the planes of the lattice.
This results in a friction force which damps the motion.

Let us therefore enquire about the relative number of atoms in the superfluid
component of a BEC in a state of motion at a given velocity $v$, $N_s(v)/N$
say, with $N$ being the number of atoms inside the harmonic trap in
the absence of the optical potential. 
In order to measure this function and to deduce a maximum 
critical velocity $v_{max}$, we have varied the displacement $\Delta x$ and
recorded the atomic distributions after a fixed evolution time
$t_{ev}=40\,$ms. For low velocities, up to about $2\,$mm/s,  
the sample follows the position of a freely oscillating BEC, 
the ratio $N_s/N$ being a constant approximately equal to $0.7$. This reduction
below unity is merely due to the loss of atoms by spontaneous scattering 
of photons from the optical lattice beams during the preparation of the BEC. 

\begin{figure}
   \begin{center}
   \parbox{6.5cm}{
   \epsfxsize 6.5cm
   \epsfbox{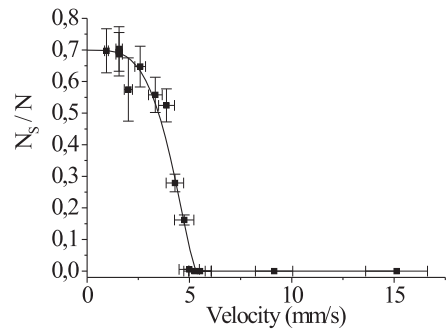}
}
   \end{center}
\caption {\small
The fraction of atoms remaining in the undistorted part of the 
BEC, $N_s/N$, as a function of 
the velocity reached during the evolution in the periodic potential.
The line shows a fit to the data assuming a 3D parabolic
density distribution and a critical velocity proportional to~$\sqrt{n(r)}$.
}
%Data from 14.12.2000
\label{N_super}
\end{figure}

Upon increasing the velocity of the BEC, we observe a retardation of a part of
the cloud, leading to a well detectable separation 
from the superfluid component after free
evolution. For velocities $v\sim 4\,$mm/s we observe that
only the central part of the fluid is moving without
retardation, leading to a drastically changed aspect ratio with respect to the
``unperturbed'' BEC. The spatial separation 
from the thermal component allows a clear demonstration 
of the superfluid properties of inhomogeneous Bose-Einstein
condensates and a precise measurement of the critical
velocity. The data for $N_s(v)/N$ in Fig.~\ref{N_super} show 
a dramatic depletion of the number of atoms in the superfluid component 
as the velocity of the fluid increases. 

To model the breakdown of superfluidity in the inhomogeneous
density distribution of the trapped BEC
in Fig.~\ref{N_super}, we first discuss the 
position dependence of the longitudinal sound velocity in the sample.
In an inhomogeneous system, as already pointed out by Andrews {\it et al.} 
~\cite{Ketterle1997a} in discussing sound propagation in a magnetically
trapped BEC, one may define a local speed of sound 
$c_s(r)$. This is given by 
$c_s(r)=\sqrt{(n(r)/m)(\delta\mu/\delta n) }$~\cite{BP}, where
within the Bogolubov approximation
the stiffness constant $\delta\mu/\delta n$ would be equal to 
the coupling strength $g$.
We have evaluated the appropriate stiffness constant by giving a longitudinal 
stretching (or squeezing) to our model system by a relative amount 
$\epsilon$ of order $1\%$ 
at fixed transverse profile. We find $\delta\mu/(\mu\epsilon)
\simeq 0.7$. Taking $\mu/k_B\simeq 170\,$nK and 
$\delta n\simeq -n\epsilon$ where 
$n$ is the average density ($n\simeq 0.4\,n_{max}\simeq 
6\times 10^{13}\,$cm$^{-3}$), we find $c_{s,max}\simeq 5.2\,$mm/s for the
maximum value of the local sound velocity at the peak of the BEC density.
This is in excellent agreement with the data in Fig.~\ref{N_super}, 
showing that complete destruction of the superfluid component occurs 
at $v_{max}\simeq 5\,$mm/s.

Assuming, therefore, that the critical velocity $v_{c}(r)$ 
for local destruction of the
superfluid component in the inhomogeneous condensate coincides with the 
local speed of sound $c_s(r)$, we have  
$v_c(r) \propto \sqrt{n(r)}$. 
As observed, superfluidity breaks down first in the
low-density regions. The envelope function of the density
distribution of the BEC is an inverted parabola in 3D (see
Fig.~\ref{simulation_groundstate}) and hence,  by 
integration over the high-density region,  we get an equation for
the relative number of atoms in the superfluid part of the BEC
for a given velocity~$v$,
$N_s(v)/N=\left[5/2\times(1-v^2/v_{max}^2)^{3/2}-3/2\times
(1-v^2/v_{max}^2)^{5/2}\right]$.
This expression implies that about 90\% of the atomic probability
density is localized in a region which remains superfluid up to
velocities $v \simeq v_{max}/2$.
The line in Fig.~\ref{N_super} shows that the above expression for 
$N_s(v)/N$ gives a very good account of the data, the fitted 
value of the maximum velocity being $v_{max}= (5.3 \pm 0.5)\,$mm/s.

In further experiments we have also observed indications 
that the dissipation onset occurs at higher velocities for decreasing
$V_0$ and that the BEC propagates without dissipation in a regime of
very low atom number. We plan to  investigate these behaviors
in detail in future work.

In conclusion, we have investigated the dynamics of BEC's in a
periodically modulated potential, both experimentally and in
numerical simulations. By measuring the effect of the periodic potential on
the sloshing-mode oscillation inside the harmonic trap we have determined an
average effective mass of the atoms in the condensate. The combined use 
of a periodic optical potential with the harmonic confinement has allowed us to
observe novel features of superfluidity in an inhomogeneous atomic BEC
and to demonstrate a new technique for measuring a local 
density-dependent critical velocity.

The results of this work are of importance for future experiments
using periodic potentials for the manipulation of Bose-Einstein
condensates and for the understanding of dissipative processes in
coherent matter waves.
The precise control of the parameters promises to be a powerful
tool for a quantitative exploration of novel regimes
occuring at different atom numbers or tunnel barrier heights.

We acknowledge discussions with M.\,Artoni and
A.\,Smerzi and support by  the EU under Contract
HPRI-CT 1999-00111 and HPRN-CT-2000-00125
and by MURST through  the PRIN\,1999 and PRIN\,2000  Initiatives.

\end{document}